\documentclass{article}
\usepackage{amsmath}
\usepackage{hyperref}
\usepackage{cleveref}
\usepackage{dieudonne}
\usepackage{macros}
\usepackage{jon-amsthm}
\usepackage{jon-biblatex}
\usepackage{newtxtext}
\usepackage{newtxmath}

\hypersetup{
  breaklinks=true,
  colorlinks=true,
  linkcolor=black,
  urlcolor=Matterhorn,
  citecolor=Matterhorn,
  linktocpage=false
}

\RenewDocumentCommand\DieudonneInterpunct{}{\textasteriskcentered}
\usepackage[outermarks]{titlesec}
\titleformat{\section}
  {\normalfont\Large\bfseries}{\thesection}{1em}{}
\titleformat{\subsection}
  {\normalfont\large\bfseries}{\thesubsection}{1em}{}
\titleformat{\subsubsection}
  {\normalfont\normalsize\bfseries}{\thesubsubsection}{1em}{}
\titleformat{\paragraph}
  {\normalfont\normalsize\bfseries}{\theparagraph}{1em}{}
\titleformat{\subparagraph}
  {\normalfont\normalsize\bfseries}{\thesubparagraph}{1em}{}

\titlespacing*{\section}      {0pt}{3.5ex plus 1ex minus .2ex} {2.3ex plus .2ex}
\titlespacing*{\subsection}   {0pt}{3.25ex plus 1ex minus .2ex}{1.5ex plus .2ex}
\titlespacing*{\subsubsection}{0pt}{3.25ex plus 1ex minus .2ex}{1.5ex plus .2ex}
\titlespacing*{\paragraph}    {0pt}{3.25ex plus 1ex minus .2ex}{1.5ex plus .2ex}
\titlespacing*{\subparagraph} {0pt}{3.25ex plus 1ex minus .2ex}{1.5ex plus .2ex}

\tikzset{
  emb/.style = {
    {right hook}-{Triangle[open]}
  },
  proj/.style = {
    -{Triangle[open]}
  },
  semb/.style = {
    >>/.tip = {.Circle[open]Triangle[open]},{right hook}->>
  },
  sproj/.style = {
    >>/.tip = {.Circle[open]Triangle[open]},->>
  },
  strict/.style = {
    -{Circle[open]},
  },
}

\let\mathscr\mathcal

\title{Bilimits in categories of partial maps}
\author{Jonathan Sterling}
\date{\today}

\NewDocumentCommand\Lift{}{\Con{L}}
\NewDocumentCommand\dcpo{}{\Con{dcpo}}
\NewDocumentCommand\pdcpo{}{\Con{pdcpo}}
\NewDocumentCommand\dcppo{}{\Con{dcppo}}
\NewDocumentCommand\EP{m}{\DelimMin{1}#1\Sup{\Con{E}}}
\NewDocumentCommand\PE{m}{\DelimMin{1}#1\Sup{\Con{P}}}

\ExplSyntaxOn
\NewDocumentCommand\PMor{D||{} o m m}{
  \jms_tikz_auto_morphism:nnnn {-left~to,#1} {#2} {#3} {#4}
}
\NewDocumentCommand\SMor{D||{} o m m}{
  \jms_tikz_auto_morphism:nnnn {strict,#1} {#2} {#3} {#4}
}
\RenewDocumentCommand\EmbMor{D||{} o m m}{
  \jms_tikz_auto_morphism:nnnn {emb,#1} {#2} {#3} {#4}
}
\NewDocumentCommand\ProjMor{D||{} o m m}{
  \jms_tikz_auto_morphism:nnnn {proj,#1} {#2} {#3} {#4}
}
\NewDocumentCommand\SEmbMor{D||{} o m m}{
  \jms_tikz_auto_morphism:nnnn {semb,#1} {#2} {#3} {#4}
}
\NewDocumentCommand\SProjMor{D||{} o m m}{
  \jms_tikz_auto_morphism:nnnn {sproj,#1} {#2} {#3} {#4}
}
\ExplSyntaxOff

\begin{document}

\maketitle

\begin{abstract}
  The closure of chains of embedding-projection pairs (\emph{ep-pairs}) under
  bilimits in some categories of predomains and domains is standard and
  well-known. For instance, Scott's $D_\infty$ construction is well-known to
  produce directed bilimits of ep-pairs in the category of directed-complete
  partial orders, and de Jong and Escard\'o have formalized this result in the
  constructive domain theory of a topos. The explicit construcition of bilimits
  for categories of predomains and \emph{partial} maps is considerably murkier
  as far as constructivity is concerned; most expositions employ the
  constructive taboo that every \emph{lift}-algebra is free, reducing the
  problem to the construction of bilimits in a category of pointed domains and
  strict maps.  An explicit construction of the bilimit is proposed in the
  dissertation of Claire Jones, but no proof is given so it remained unclear if
  the category of dcpos and partial maps was closed under directed bilimits of
  ep-pairs in a topos. We provide a (Grothendieck)-topos-valid proof that the
  category of dcpos and partial maps between them is closed under bilimits;
  then we describe some applications toward models of axiomatic and synthetic
  domain theory.
\end{abstract}

\tableofcontents

\bigskip

\begin{node}
  \emph{Acknowledgment.}
  Thanks to Lars Birkedal, Mart\'in Escard\'o, Marcelo Fiore, Daniel Gratzer, and Tom de Jong
  for their assistance while preparing this note.
\end{node}

\section{Preliminaries}

\begin{node}
  In an poset-enriched category $\ECat$, we define an \emph{embedding} $\EmbMor{U}{A}$ to be
  a monomorphism $\Mor|>->|[\epsilon]{U}{A}$ that has
  a right adjoint, called its \emph{projection} $\Mor[\pi]{A}{U}$. Because
  $\epsilon\dashv\pi$ and $\epsilon$ is mono, we have $\pi\circ\epsilon=\ArrId{U}$ and
  $\epsilon\circ\pi\leq\ArrId{A}$.
\end{node}

\begin{node}
  Dually we define a \emph{projection} $ \ProjMor{A}{U}$ to be a map
  $\Mor|->>|[\epsilon]{A}{U}$ that has a monomorphic left adjoint. Embeddings
  and projections are uniquely determined by each other.
\end{node}

\begin{node}
  We will write $\EP{\ECat}$ for the wide subcategory of $\ECat$ spanned by
  embeddings, and $\PE{\ECat}$ for the wide subcategory of $\ECat$ spanned by
  projections. We note that $\PE{\ECat} = \OpCat{\prn{\EP{\ECat}}}$.
\end{node}

\NewDocumentCommand\ICat{}{\CatIdent{I}}

\section{Bilimits of directed diagrams in \texorpdfstring{$\EP{\dcpo}$}{dcpoE}}

\begin{node}
  Let $\ICat$ be a directed poset, \ie a filtered category whose hom sets are
  propositions. \citet{dejong-escardo:2021} have verified constructively that
  $\PE{\dcpo}$ is closed under limits of $\OpCat{\ICat}$-diagrams, and that, moreover,
  the cocone obtained from the embeddings of the universal cone is colimiting.
\end{node}

\subsection{Limits of co-directed diagrams}

\begin{node}
  The limit node of a diagram $\Mor[D_\bullet]{\OpCat{\ICat}}{\PE{\dcpo}}$ is explicitly
  explicitly computed as the following dcpo equipped with the pointwise order:
  \[
    D_\infty \coloneqq \Compr{\sigma : \Prod{i\in\ICat}D_i}{\forall i\leq j\in\ICat. \pi\Sub{i\leq j}\sigma_j = \sigma_i}
  \]
\end{node}

\subsubsection{Constructing the limiting cone}

\begin{node}\label{node:tot:limit-cone}
  The universal cone $\ProjMor{\brc{D_\infty}}{D_\bullet}$ in $\brk{\OpCat{\ICat},\PE{\dcpo}}$ is defined like so:
  \begin{align*}
    \pi\Sub{i<\infty} &: \ProjMor{D_\infty}{D_i}\\
    \pi\Sub{i<\infty} \sigma &= \sigma_i
  \end{align*}

  We verify that $\pi\Sub{i<\infty}$ is a projection by defining its left adjoint explicitly:
  \begin{align*}
    \epsilon\Sub{i<\infty} &: \EmbMor{D_i}{D_\infty}\\
    \prn{\epsilon\Sub{i<\infty} x}_j &= \pi\Sub{j\leq k}\epsilon\Sub{i \leq k} x\quad\text{for $k\geq i,j$}
  \end{align*}

  Diagrammatically, each component of $\epsilon\Sub{i<\infty}$ is given like so:
  \[
    \begin{tikzpicture}
      \node (i) {$D_i$};
      \node (k) [right = of i] {$D_k$};
      \node (j) [right = of k] {$D_j$};
      \draw[emb] (i) to (k);
      \draw[proj] (k) to (j);
    \end{tikzpicture}
  \]

  We check diagrammatically that $\epsilon\Sub{i<\infty}$ in fact takes values in
  $D_\infty$:
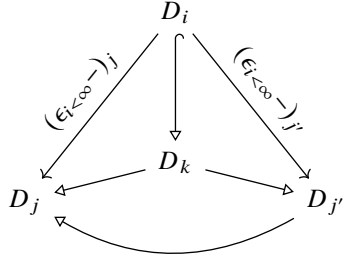
  \[
    \begin{tikzpicture}[diagram]
      \node (i) {$D_i$};
      \node (k) [below = of i] {$D_k$};
      \node (j') [right = of k,yshift=-.5cm] {$D_{j'}$};
      \node (j) [left = of k,yshift=-.5cm] {$D_j$};
      \draw[emb] (i) to (k);
      \draw[proj] (k) to (j);
      \draw[->] (i) to node [sloped,above] {$\prn{\epsilon\Sub{i<\infty}-}_j$} (j);
      \draw[->] (i) to node [sloped,above] {$\prn{\epsilon\Sub{i<\infty}-}_{j'}$} (j');
      \draw[proj] (k) to (j');
      \draw[proj,bend left=30] (j') to (j);
    \end{tikzpicture}
  \]

  We easily verify that $\epsilon\Sub{i<\infty}$ is a section of
  $\pi\Sub{i<\infty}$. It remains to check that $\epsilon\Sub{i<\infty}\circ
  \pi\Sub{i<\infty}\leq \ArrId{D_\infty}$; because the order is pointwise on
  $D_\infty$ it actually suffices to check that $\pi\Sub{j<\infty}\circ
  \epsilon\Sub{i<\infty}\circ \pi\Sub{i<\infty} \leq \pi\Sub{j<\infty}$ for
  each $j\in\ICat$. Fixing $\sigma\in D_\infty$ we may compute:
  \[
    \pi\Sub{j<\infty}\epsilon\Sub{i<\infty}\pi\Sub{i<\infty} \sigma
    =
    \pi\Sub{j\leq k}\epsilon\Sub{i\leq k}\sigma_i
    \leq
    \sigma_i
  \]

  The right-hand inequality holds because each $\epsilon\Sub{i\leq k}\dashv
  \pi\Sub{i\leq k}$ is an ep-pair.
\end{node}

\begin{node}
  We elaborate on the fact that the embedding
  $\EmbMor[\epsilon\Sub{i<\infty}]{D_i}{D_\infty}$ is well-defined: at least
  one such $k\geq i,j$ must exist because $\ICat$ is directed, but we must also
  argue that the definition does not depend on the particular choice of $k$.
  Indeed, suppose that we choose two different $k,k'$ as in the following
  scenario:
  
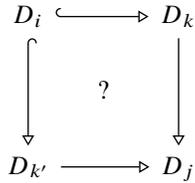
\[
    \begin{tikzpicture}[diagram]
      \SpliceDiagramSquare{
        north/style = emb,
        east/style = proj,
        south/style = proj,
        west/style = emb,
        nw = D_i,
        ne = D_k,
        se = D_j,
        sw = D_{k'}
      }
      \node [between = nw and se] {$?$};
    \end{tikzpicture}
  \]

  There exists $m\geq k,k'$ which we use to verify that the diagram commutes:
  
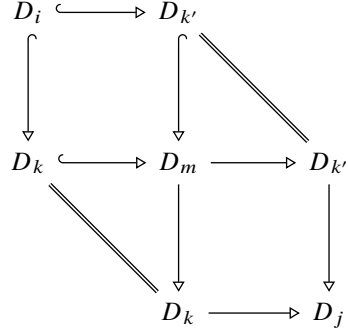
\[
    \begin{tikzpicture}[diagram]
      \node (i) {$D_i$};
      \node (k/1) [below = of i] {$D_k$};
      \node (m) [right = of k/1] {$D_m$};
      \node (k/2) [below = of m] {$D_k$};
      \node (j) [right = of k/2] {$D_j$};
      \node (k'/1) [above = of m] {$D_{k'}$};
      \node (k'/2) [right = of m] {$D_{k'}$};
      \draw[emb] (i) to (k/1);
      \draw[emb] (i) to (k'/1);
      \draw[emb] (k/1) to (m);
      \draw[emb] (k'/1) to (m);
      \draw[proj] (m) to (k/2);
      \draw[proj] (k/2) to (j);
      \draw[proj] (m) to (k'/2);
      \draw[proj] (k'/2) to (j);
      \draw[double] (k/1) to (k/2);
      \draw[double] (k'/1) to (k'/2);
    \end{tikzpicture}
  \]
\end{node}

\begin{node}\label{node:lub-of-projections}
  Any element $\sigma \in D_\infty$ is the least upper bound of its family of
  approximations $\brc{\epsilon\Sub{i<\infty}\pi\Sub{i<\infty}\sigma \leq
  \sigma}$.  Indeed, fix any $\sigma'\in D_\infty$ greater than each
  $\epsilon\Sub{i<\infty}\pi\Sub{i<\infty}\sigma$; we will verify that
  $\sigma\leq \sigma'$. Because the order on $D_\infty$ is pointwise, it
  suffices to check that for each $j\in\ICat$, we have $\pi\Sub{j<\infty}\sigma
  \leq \pi\Sub{j<\infty}\sigma'$.
  Unfolding our assumption, for any $i,j\in\ICat$ and $k\geq i,j$ we have
  $\pi\Sub{j\leq k}\epsilon\Sub{i\leq k} \pi\Sub{i<\infty}\sigma \leq
  \pi\Sub{j<\infty} \sigma'$. Setting $i=j=k$, our goal follows.
\end{node}

\subsubsection{Universal property of the limiting cone}

\begin{node}
  Fix another cone $\ProjMor[p_\bullet]{\brc{H}}{D_\bullet}$ in $\brk{\OpCat{\ICat},\PE{\dcpo}}$; we will
  exhibit the unique projection $\ProjMor[p_\infty]{H}{D_\infty}$ making the following
  commute:
  
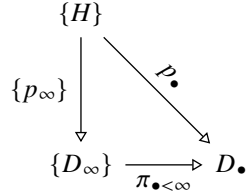
\[
    \begin{tikzpicture}[diagram]
      \node (D/w) {$\brc{D_\infty}$};
      \node (D) [right = of D/w] {$D_\bullet$};
      \node (H) [above = of D/w] {$\brc{H}$};
      \draw[proj] (D/w) to node [below] {$\pi\Sub{\bullet<\infty}$} (D);
      \draw[proj] (H) to node [left] {$\brc{p_\infty}$} (D/w);
      \draw[proj] (H) to node [sloped,above] {$p_\bullet$} (D);
    \end{tikzpicture}
  \]

  Note that by duality, $h$ is equivalently a \emph{cocone} in $\brk{\ICat,\EP{\dcpo}}$.
\end{node}

\paragraph{Constructing the mediating map}

\begin{node}\label{node:total:mediating-map:defi}
  To define $\ProjMor[p_\infty]{H}{D_\infty}$ we set each
  $\Mor[\prn{p_\infty}_i]{H}{D_i}$ to be simply $p_i$; that this determines an
  element of $D_\infty$ is exactly the naturality of
  $\ProjMor[p_\bullet]{\brc{H}}{D_\bullet}$. To see that $p_\infty$
  so-described is a projection, we will explicitly construct its left adjoint
  $e_\infty\dashv p_\infty$. To define the embedding
  $\EmbMor[e_\infty]{D_\infty}{H}$, we will use all of the embeddings $e_i\dashv p_i$:
  \begin{align*}
    e_\infty &: \EmbMor{D_\infty}{H}\\
    e_\infty &= \DLub{i\in \ICat} e_i \circ \pi\Sub{i<\infty}
  \end{align*}

  To illustrate, we are taking the least upper bound of the following $\ICat$-indexed set of maps:
  \[
    \begin{tikzpicture}[diagram]
      \node (D/w) {$D_\infty$};
      \node (D/i) [right = of D/w] {$D_i$};
      \node (H) [right = of D/i] {$H$};
      \draw[proj] (D/w) to node [below] {$\pi\Sub{i<\infty}$} (D/i);
      \draw[emb] (D/i) to node [below] {$e_i$} (H);
    \end{tikzpicture}
  \]
\end{node}

\begin{node}
  We note that the mediating map $\ProjMor[p_\infty]{H}{D_\infty}$ commutes
  with the projections by definition, \ie we need $\pi\Sub{i<\infty}\circ
  p_\infty = p_i$ for each $i\in \ICat$.
\end{node}

\paragraph{The mediating map is a projection}

\begin{node}\label{node:total:mediating-map-is-retraction}
  We need to check that $e_\infty$ is a section of $p_\infty$ so-defined; it
  suffices to check that each of the triangles below commutes, since $D_\infty$
  is a subobject of the product $\Prod{i\in\ICat}D_i$:
  \[
    \begin{tikzpicture}[diagram]
      \node (D/w) {$D_\infty$};
      \node (H) [right = of D/w] {$H$};
      \node (D/i) [below = of H] {$D_i$};
      \draw[emb] (D/w) to node [above] {$e_\infty$} (H);
      \draw[proj] (H) to node [right] {$p_i$} (D/i);
      \draw[proj] (D/w) to node [sloped,below] {$\pi\Sub{i<\infty}$} (D/i);
    \end{tikzpicture}
  \]

  By definition of $e_\infty$, the right-hand composite
  $p_i\circ e_\infty$ is the least upper bound of
  the following directed family of maps indexed in $j\in\ICat$, so it suffices
  to check that $\pi\Sub{i<\infty}$ is \emph{also} the
  upper bound of the same:
  \[
    \begin{tikzpicture}[diagram]
      \node (D/w) {$D_\infty$};
      \node (D/j) [right = of D/w] {$D_j$};
      \node (H) [right = of D/j] {$H$};
      \node (D/i) [right = of H] {$D_i$};
      \draw[proj] (D/w) to node [above] {$\pi\Sub{j<\infty}$} (D/j);
      \draw[emb] (D/j) to node [above] {$e_j$} (H);
      \draw[proj] (H) to node [above] {$p_i$} (D/i);
    \end{tikzpicture}
  \]

  For arbitrary $k\geq i,j$ we may factor out $H$ from the above composite:
  \[
    \begin{tikzpicture}[diagram]
      \node (D/w) {$D_\infty$};
      \node (D/j) [right = of D/w] {$D_j$};
      \node (H) [right = of D/j] {$H$};
      \node (D/i) [right = of H] {$D_i$};
      \draw[proj] (D/w) to node [above] {$\pi\Sub{j<\infty}$} (D/j);
      \draw[emb] (D/j) to node [above] {$e_j$} (H);
      \draw[proj] (H) to node [above] {$p_i$} (D/i);
      \node (D/k) [between = D/j and H, yshift=-1.5cm] {$D_k$};
      \node (D/k') [between = H and D/i, yshift=-1.5cm] {$D_k$};
      \draw[double] (D/k) to (D/k');
      \draw[emb] (D/j) to node [sloped,below] {$\epsilon\Sub{j\leq k}$} (D/k);
      \draw[emb] (D/k) to node [desc] {$e_k$} (H);
      \draw[proj] (H) to node [desc] {$p_k$} (D/k');
      \draw[proj] (D/k') to node [sloped,below] {$\pi\Sub{i<k}$} (D/i);
    \end{tikzpicture}
  \]

  The above is evidently equal to the following composite:
  \[
    \begin{tikzpicture}[diagram]
      \node (D/w) {$D_\infty$};
      \node (D/j) [right = of D/w] {$D_j$};
      \node (D/w') [right = of D/j] {$D_\infty$};
      \node (D/i) [right = of H] {$D_i$};
      \draw[proj] (D/w) to node [above] {$\pi\Sub{j<\infty}$} (D/j);
      \draw[emb] (D/j) to node [above] {$\epsilon\Sub{j<\infty}$} (D/w');
      \draw[proj] (H) to node [above] {$\pi\Sub{i<\infty}$} (D/i);
    \end{tikzpicture}
  \]

  By \cref{node:lub-of-projections} we have $\ArrId{D_\infty} =
  \DLub{j\in\ICat}\epsilon\Sub{j<\infty}\circ\pi\Sub{j<\infty}$, so we
  immediately have $\pi\Sub{i<\infty} =
  \DLub{j\in\ICat}\pi\Sub{i<\infty}\circ\epsilon\Sub{j<\infty}\circ\pi\Sub{j<\infty}$
  as desired. Therefore $e_\infty$ is a section of $p_\infty$.
\end{node}

\begin{node}
  It remains to check that $e_\infty\circ p_\infty \leq \ArrId{H}$; the
  universal property of $\EmbMor[e_\infty]{D_\infty}{H}$ as a least upper bound
  is defined ensures that it suffices to check that each of the following
  composites is smaller than the identity, which follows from our assumption
  that $e_i\dashv p_i$ is an embedding-projection pair:
  \[
    \begin{tikzpicture}[diagram]
      \node (H) {$H$};
      \node (D/w) [right = of H] {$D_\infty$};
      \node (D/i) [right = of D/w] {$D_i$};
      \node (H') [right = of D/i] {$H$};
      \draw[proj] (H) to node [above] {$p_\infty$} (D/w);
      \draw[proj] (D/w) to node [above] {$\pi\Sub{i<\infty}$} (D/i);
      \draw[proj,bend right=30] (H) to node [below] {$p_i$} (D/i);
      \draw[emb] (D/i) to node [below] {$e_i$} (H');
    \end{tikzpicture}
  \]
\end{node}

\paragraph{Uniqueness of the mediating map}
We must argue that $\ProjMor[p_\infty]{H}{D_\infty}$ is the \emph{only} projection map making
the following diagram commute:
\[
  \begin{tikzpicture}[diagram]
    \node (D/w) {$\brc{D_\infty}$};
    \node (D) [right = of D/w] {$D_\bullet$};
    \node (H) [above = of D/w] {$\brc{H}$};
    \draw[proj] (D/w) to node [below] {$\pi\Sub{\bullet<\infty}$} (D);
    \draw[proj] (H) to node [left] {$\brc{p_\infty}$} (D/w);
    \draw[proj] (H) to node [sloped,above] {$p_\bullet$} (D);
  \end{tikzpicture}
\]

This is easily deduced at level of points by considering the universal
property of the product $\Prod{i\in\ICat}D_i$ of which $D_\infty$ is a
subposet.

\subsection{As a directed colimit of embeddings}

\begin{node}
  As the identification $\EP{\dcpo} = \OpCat{\prn{\PE{\dcpo}}}$ proceeds by
  swapping embeddings for projections, we see that the embeddings corresponding
  to the limit cone for the diagram
  $\Mor[D_\bullet]{\OpCat{\ICat}}{\PE{\dcpo}}$ induce a colimiting cone for the
  equivalent diagram $\Mor[D_\bullet]{\ICat}{\EP{\dcpo}}$.
\end{node}

\begin{node}
  Therefore we have constructed what some refer to as a \emph{bilimit} in the
  category of embedding-projection pairs over $\dcpo$.
\end{node}

\section{Bilimits of directed diagrams in $\EP{\pdcpo}$}

\subsection{Lifting and partial maps}

\begin{node}
  The existence of bilimits of directed diagrams in $\EP{\pdcpo}$ is folklore;
  the result easily follows from more general considerations when $\pdcpo =
  \dcppo$, but this identification relies on classical logic. Under slightly
  different assumptions, a proof is sketched by \citet{fiore:1994}, and an
  elementary construction of the bilimit is claimed and not proved by
  \citet{jones:1990,jones-plotkin:1989}.  We will generalize the scenario
  discussed by \opcit{} to speak of limits of $\OpCat{\ICat}$-indexed diagrams of
  projections in $\PE{\pdcpo}$.
\end{node}

\begin{node}
  The \emph{lift} monad $\Mor[\Lift]{\dcpo}{\dcpo}$ must be defined more
  carefully than in a classical setting. In particular, we set the carrier set
  of $\Lift{A}$ to be the \emph{partial map classifier} of $A$:
  \[
    \Lift{A} = \Sum{\phi:\Omega}{\prn{\phi\Rightarrow A}}
  \]

  We will write $\IsDefd{u}$ to mean $\pi_1\,u = \top$; we treat the second
  projection implicitly in most cases.
  We impose the order $u\leq v \Leftrightarrow \forall z : \IsDefd{u}. \exists
  z' :\IsDefd{v}. u\,z \leq v\,z'$; in the future we will be less precise and
  write things like $\IsDefd{u}\Rightarrow \prn{\IsDefd{v} \land u \leq v}$ to
  mean the same thing.
  Given a directed family of elements $\Compr{u_i\in\Lift{A}}{i\in \ICat}$, the
  least upper bound $\DLub{i\in\ICat}{u_i}$ is defined to be the least upper
  bound in $A$ of the directed family $\Compr{u_i\in A}{i\in\ICat \text{ s.t. }
  \IsDefd{u_i}}$.
\end{node}

\begin{node}
  To simplify matters, we note that although partial map $\PMor{A}{B}$ is
  defined to be an ordinary map $\Mor[f]{A}{\Lift{B}}$, we may equivalently
  define it in terms of the \emph{strict} map $\SMor[f^\sharp]{\Lift{A}}{\Lift{B}}$. In
  other words, the Kleisli category $\pdcpo$ can be identified with the full
  subcategory of $\dcppo$ spanned by \emph{free} $\Lift$-algebras, \ie objects
  of the form $\Lift{A}$. In our presentation, we will say that an object of
  $\pdcpo$ is a dcpo and a morphism from $A$ to $B$ is a strict map
  $\SMor{\Lift{A}}{\Lift{B}}$. The advantage of this presentation of $\pdcpo$
  is that composition of maps is given as in $\dcpo$ rather than via the
  Kleisli extension.
\end{node}

\subsection{Limits of co-directed diagrams of partial projections}

\begin{node}
  Let $\Mor[D_\bullet]{\OpCat{\ICat}}{\PE{\dcpo}}$ be a diagram of partial
  projections, \ie for each $i\leq j$ we have a strict map
  $\SProjMor[\pi\Sub{i\leq j}]{\Lift{D_j}}{\Lift{D_i}}$ such that there exists
  $\SEmbMor[\epsilon\Sub{i\leq j}]{\Lift{D_i}}{\Lift{D_j}}$ with $\pi\Sub{i\leq
  j}\circ \epsilon\Sub{i\leq j} = \ArrId{\Lift{D_i}}$ and $\epsilon\Sub{i\leq
  j}\circ\pi\Sub{i\leq j} \leq \ArrId{\Lift{D_j}}$. Following \citet{jones-plotkin:1989} we define
  the limit node $D_\infty$ to be the following dcpo equipped with the pointwise order:
  \[
    D_\infty =
    \Compr{
      \sigma:\Prod{i\in\ICat} \Lift{D_i}
    }{
      \prn{\exists i\in\ICat.\IsDefd{\sigma_i}}
      \land
      \forall i\leq j\in\ICat.
      \pi\Sub{i\leq j}\sigma_j = \sigma_i
    }
  \]
\end{node}

\subsubsection{Constructing the limiting cone}

\begin{node}
  The universal cone $\SProjMor{\brc{\Lift{D_\infty}}}{\Lift{D_\bullet}}$ in
  $\brk{\OpCat{\ICat},\PE{\pdcpo}}$ is defined like so:
  \begin{align*}
    \pi\Sub{i<\infty} &: \SProjMor{\Lift{D_\infty}}{\Lift{D_i}}\\
    \pi\Sub{i<\infty} &= \mu\Sub{D_i} \circ \Lift{\pi_i}
  \end{align*}
  where $\Mor[\pi_i]{\Prod{j\in\ICat}{\Lift{D_j}}}{\Lift{D_i}}$ is the obvious
  map in $\dcpo$ and $\SMor[\mu_{D}]{\Lift^2{D}}{\Lift{D}}$ is the
  multiplication map for the lift monad. Alternatively we could have written
  $\pi\Sub{i<\infty} = \pi_i^\sharp$ in terms of Kleisli extension.

  To see that $\pi\Sub{i<\infty}$ is a projection, we construct the
  corresponding embedding.
  \begin{align*}
    \epsilon\Sub{i<\infty} &: \SEmbMor{\Lift{D_i}}{\Lift{D_\infty}}\\
    \epsilon\Sub{i<\infty}x &= \Lift\,{\brk{j\mapsto \pi\Sub{j\leq k}\circ \epsilon\Sub{i\leq k}\circ\eta\Sub{D_i}}} && \text{for $k\geq i,j$}
  \end{align*}

  Above we have written $\brk{j\mapsto \pi\Sub{j\leq k}\circ\epsilon\Sub{i\leq
  k}}$ for the map $\Mor{D_i}{D_\infty}$ determined by the following
  product cone indexed in $j\in\ICat$:
  \[
    \begin{tikzpicture}[diagram]
      \node (i') {$D_i$};
      \node (i) [right = of i'] {$\Lift{D_i}$};
      \node (k) [right = of i] {$\Lift{D_k}$};
      \node (j) [right = of k] {$\Lift{D_j}$};
      \draw[>->] (i') to (i);
      \draw[semb] (i) to (k);
      \draw[sproj] (k) to (j);
    \end{tikzpicture}
  \]

  It is clear that $\epsilon\Sub{i<\infty}$ in fact takes values in
  $\Lift{D_\infty}$ because the result is at least defined at level $i$.  We
  must check that $\epsilon\Sub{i<\infty}$ is a section of $\pi\Sub{i<\infty}$;
  considering the uniqueness of $\Lift$-extensions, it suffices to check that
  the following commutes:
  \[
    \begin{tikzpicture}[diagram]
      \node (D/i) {$D_i$};
      \node (L/D/i) [right = of D/i] {$\Lift{D_i}$};
      \node (L/D/w) [right = of L/D/i] {$\Lift{D_\infty}$};
      \node (L/D/i') [below = of L/D/w] {$\Lift{D_i}$};
      \draw[>->] (D/i) to node [above] {$\eta\Sub{D_i}$} (L/D/i);
      \draw[>->] (D/i) to node [sloped,below] {$\eta\Sub{D_i}$} (L/D/i');
      \draw[semb] (L/D/i) to node [above] {$\epsilon\Sub{i<\infty}$} (L/D/w);
      \draw[sproj] (L/D/w) to node [right] {$\pi\Sub{i<\infty}$} (L/D/i');
    \end{tikzpicture}
  \]

  We verify the above using the fact that each $\epsilon\Sub{i\leq k}\dashv\pi\Sub{i\leq k}$ is an ep-pair:
  \begin{align*}
    \pi\Sub{i<\infty}\circ\epsilon\Sub{i<\infty}\circ\eta\Sub{D_i} &=
    \pi\Sub{i<\infty}\circ\eta\Sub{D_\infty}\circ\prn{j\mapsto \pi\Sub{j\leq k}\circ\epsilon\Sub{i\leq k}\circ \eta\Sub{D_i}}
    \\
    &=
    \pi\Sub{i\leq k}\circ\epsilon\Sub{i\leq k}\circ\eta\Sub{D_i}
    \\
    &=
    \eta\Sub{D_i}
  \end{align*}

  We check that $\epsilon\Sub{i<\infty}\circ\pi\Sub{i<\infty} \leq
  \ArrId{\Lift{D_\infty}}$ in the same way as for the total case.
\end{node}

\begin{node}
  Any element $\sigma\in \Lift{D_\infty}$ is the least
  upper bound of its family of approximations
  $\brc{\epsilon\Sub{i<\infty}\pi\Sub{i<\infty}\sigma \leq \sigma}$.  Fix any
  other $\sigma'\in \Lift{D_\infty}$ greater than each
  $\epsilon\Sub{i<\infty}\pi\Sub{i<\infty}\sigma$ to check that $\sigma\leq
  \sigma'$.
  Assuming $\sigma = \eta\Sub{D_\infty}\tau$ we must check that there exists
  some $\tau'\in D_\infty$ such that $\sigma'=\eta\Sub{D_\infty}\tau'$ and
  $\tau\leq\tau'$. Because at least one projection of $\tau$ is defined and
  $\sigma'$ is greater than every projection of $\tau$, it must be that
  $\sigma'$ is defined; therefore, we may fix $\tau'$ with
  $\sigma'=\eta\Sub{D_\infty}\tau'$ and proceed to check that $\tau\leq \tau'$,
  which follows \emph{mutatis mutandis} as in the total
  case~\cref{node:lub-of-projections}.
\end{node}

\subsubsection{Universal property of the limiting cone}

\begin{node}\label{node:partial:mediating-map}
  Fix another cone $\SProjMor[p_\bullet]{\Lift{H}}{\Lift{D_\bullet}}$ in
  $\brk{\OpCat{\ICat},\PE{\pdcpo}}$. We will exhibit the unique strict
  projection $\SProjMor[p_\infty]{\Lift{H}}{\Lift{D_\infty}}$ making the
  following commute:
  \[
    \begin{tikzpicture}[diagram]
      \node (D/w) {$\brc{\Lift{D_\infty}}$};
      \node (D) [right = of D/w] {$\Lift{D_\bullet}$};
      \node (H) [above = of D/w] {$\brc{\Lift{H}}$};
      \draw[sproj] (D/w) to node [below] {$\pi\Sub{\bullet<\infty}$} (D);
      \draw[sproj] (H) to node [left] {$\brc{p_\infty}$} (D/w);
      \draw[sproj] (H) to node [sloped,above] {$p_\bullet$} (D);
    \end{tikzpicture}
  \]
\end{node}

\paragraph{Constructing the mediating map}

\begin{node}
  We first define the termination support of
  $\SProjMor[p_\infty]{\Lift{H}}{\Lift{D_\infty}}$ as a strict continuous map
  $\SMor[\IsDefd{p_\infty}]{\Lift{H}}{\Omega}$, which we obtain as the least
  upper bound of the following directed family of maps indexed in $i\in\ICat$:
  \[
    \begin{tikzpicture}[diagram]
      \node (L/H) {$\Lift{H}$};
      \node (L/D/i) [right = of L/H] {$\Lift{D_i}$};
      \node (W) [right = of L/D/i] {$\Omega$};
      \draw[sproj] (L/H) to node [above] {$p_i$} (L/D/i);
      \draw[strict] (L/D/i) to node [above] {$\IsDefd{-}$} (W);
      \draw[strict,bend right=30] (L/H) to node [below] {$\IsDefd{p_i}$} (W);
    \end{tikzpicture}
  \]

  The least upper bound $\IsDefd{p_\infty} = \DLub{i\in\ICat} \IsDefd{p_i}$
  determines a subobject $\Mor|>->|{\tilde{H}}{\Lift{H}}$ by pullback along
  $\top$; we now construct a \emph{total} map
  $\Mor[\tilde{p}_\infty]{\tilde{H}}{D_\infty}$ below:
  \[
    \begin{tikzpicture}[diagram]
      \SpliceDiagramSquare{
        nw/style = pullback,
        east/style = >->,
        west/style = >->,
        west/node/style = upright desc,
        nw = \tilde{H},
        sw = \Lift{H},
        se = \Omega,
        ne = \ObjTerm{\dcpo},
        south = \IsDefd{p_\infty},
        east = \top,
        west = \prn{\IsDefd{p_\infty}}^*\top
      }
      \node (D/w) [left = 2.75cm of nw] {$D_\infty$};
      \node (prod) [left = 2.75cm of sw] {$\Prod{i\in\ICat}\Lift{D_i}$};
      \draw[->,exists] (nw) to node [above] {$\tilde{p}_\infty$} (D/w) ;
      \draw[>->] (D/w) to (prod);
      \draw[->] (sw) to node [below] {$\brk{i\mapsto p_i}$} (prod);
    \end{tikzpicture}
  \]

  To see that the factorization above is possible, we note that under
  $\IsDefd{p_\infty}$ the product cone $\brk{i\mapsto p_i}$ really does lie in
  $D_\infty$ --- because $\IsDefd{p_\infty}$ is precisely the needed assumption
  that at least \emph{one} of the $p_i$ is defined.
  We now define $p_\infty$ using the universal property of the partial map
  classifier:
  \[
    \begin{tikzpicture}[diagram]
      \SpliceDiagramSquare{
        nw/style = pullback,
        east/style = >->,
        west/style = >->,
        west/node/style = upright desc,
        south/style = {exists,strict},
        width = 3cm,
        nw = \tilde{H},
        ne = D_\infty,
        se = \Lift{D_\infty},
        sw = \Lift{H},
        north = \tilde{p}_\infty,
        east = \eta\Sub{D_\infty},
        west = \prn{\IsDefd{p_\infty}}^*\top,
        south = p_\infty,
      }
    \end{tikzpicture}
  \]

  By definition, $p_\infty$ makes the triangle in
  \cref{node:partial:mediating-map} commute; it remains to check that
  $p_\infty$ is in fact the unique \emph{projection} making that triangle
  commute.
\end{node}

\paragraph{The mediating map is a projection}

\begin{node}\label{node:partial:mediating-map:emb}
  To see that $p_\infty$ is a projection, we will explicitly construct its left
  adjoint $\SEmbMor[e_\infty]{\Lift{D_\infty}}{\Lift{H}}$ as the least upper
  bound of the following directed set of maps indexed in $i\in\ICat$, just as
  in \cref{node:total:mediating-map:defi}:
  \[
    \begin{tikzpicture}[diagram]
      \node (D/w) {$\Lift{D_\infty}$};
      \node (D/i) [right = of D/w] {$\Lift{D_i}$};
      \node (H) [right = of D/i] {$\Lift{H}$};
      \draw[sproj] (D/w) to node [above] {$\pi\Sub{i<\infty}$} (D/i);
      \draw[semb] (D/i) to node [above] {$e_i$} (H);
    \end{tikzpicture}
  \]

  In other words, we set $e_\infty =
  \DLub{i\in\ICat}{e_i\circ\pi\Sub{i<\infty}}$.
\end{node}

\begin{node}\label{node:partial:emb-i-support}
  For $i\in\ICat$, let $\Mor|>->|{\tilde{H}_i}{\Lift{H}}$ be the subobject corresponding to the
  termination support $\SMor[\IsDefd{p_i}]{\Lift{H}}{\Lift{D_i}}$ such that we
  have $\tilde{H} = \DLub{i\in\ICat}{\tilde{H}_i}$. Then for each $i\in\ICat$,
  we have a cartesian square in the following configuration:
  \[
    \DiagramSquare{
      ne = \tilde{H}_i,
      se = \Lift{H},
      sw = \Lift{D_i},
      nw = D_i,
      south = e_i,
      west = \eta\Sub{D_i},
      north = \tilde{e}_i,
      nw/style = dotted pullback,
      east/style = >->,
      west/style = >->,
      south/style = semb,
      north/style = {exists,>->},
    }
  \]

  That the above exists and is cartesian follows from the pullback lemma, using
  the fact that each $e_i$ is a section of $p_i$:
  \[
    \begin{tikzpicture}[diagram]
      \SpliceDiagramSquare<0/>{
        ne = \tilde{H}_i,
        se = \Lift{H},
        sw = \Lift{D_i},
        nw = D_i,
        south = e_i,
        west = \eta\Sub{D_i},
        north = \tilde{e}_i,
        nw/style = dotted pullback,
        east/style = >->,
        west/style = >->,
        south/style = semb,
        north/style = {exists,->},
        north/node/style = upright desc,
        south/node/style = upright desc,
        width = 2.5cm,
      }
      \SpliceDiagramSquare<1/>{
        glue = west,
        glue target = 0/,
        east/style = >->,
        south/style = sproj,
        north/node/style = upright desc,
        south/node/style = upright desc,
        ne = D_i,
        se = \Lift{D_i},
        east = \eta\Sub{D_i},
        south = p_i,
        north = \tilde{p}_i,
        width = 2.5cm,
      }
      \draw[double,bend left=30] (0/nw) to (1/ne);
      \draw[double,bend right=30] (0/sw) to (1/se);
    \end{tikzpicture}
  \]

\end{node}

\begin{node}\label{node:partial:support-of-emb-infty}
  We have a cartesian square in the following configuration:
  \[
    \DiagramSquare{
      nw/style= dotted pullback,
      east/style = >->,
      west/style = >->,
      north/style = {exists,>->},
      south/style = semb,
      nw = D_\infty,
      sw = \Lift{D_\infty},
      se = \Lift{H},
      ne = \tilde{H},
      south = e_\infty,
      north = \tilde{e}_\infty,
    }
  \]

  To see that this is the case, we see from \cref{node:partial:emb-i-support}
  that the termination support of $e_\infty$ must be the join of all the
  subobjects $\Mor|>->|{\tilde{D}_\infty^i}{\Lift{D_\infty}}$ defined below:
  \[
    \begin{tikzpicture}[diagram]
      \SpliceDiagramSquare<0/>{
        nw = \tilde{D}_\infty^i,
        ne = D_i,
        sw = \Lift{D_\infty},
        se = \Lift{D_i},
        east = \eta\Sub{D_i},
        south = \pi\Sub{i<\infty},
        north/style = {exists,->},
        nw/style = pullback,
        ne/style = pullback,
        west/style = >->,
        east/style = >->,
        south/style = sproj,
        east/node/style = upright desc,
      }
      \SpliceDiagramSquare<1/>{
        glue = west,
        glue target = 0/,
        south/style = semb,
        north/style = >->,
        ne = \tilde{H}_i,
        se = \Lift{H},
        north = \tilde{e}_i,
        south = e_i,
      }
    \end{tikzpicture}
  \]

  Observe that $\tilde{D}_\infty^i$ is the subobject spanned by total elements
  of $D_\infty$ whose $i$th projection is defined.  Therefore the join of all
  the $\tilde{D}_\infty^i$ is the subobject of $\Lift{D_\infty}$ spanned by
  total elements of $D_\infty$ that have at least one defined projection; but
  we have already ensured that any element of $D_\infty$ has at least one
  projection defined. Therefore $\DLub{i\in\ICat}{\tilde{D}_\infty^i}$  is
  $D_\infty$ itself.
\end{node}

\begin{node}\label{node:partial:tilde-emb-section}
  The total map $\tilde{e}_\infty$ is a section of $\tilde{p}_\infty$ in the
  sense that the following triangle commutes:
  \[
    \begin{tikzpicture}[diagram]
      \node (D/w) {$D_\infty$};
      \node (H') [right = of D/w] {$\tilde{H}$};
      \node (D/w') [below = of H'] {$D_\infty$};
      \draw[>->] (D/w) to node [above] {$\tilde{e}_\infty$} (H');
      \draw[->] (H') to node [right] {$\tilde{p}_\infty$} (D/w');
      \draw[double] (D/w) to (D/w');
    \end{tikzpicture}
  \]

  It suffices to check that each of the following triangles commutes:
  \[
    \begin{tikzpicture}[diagram]
      \node (D/w) {$D_\infty$};
      \node (H') [right = of D/w] {$\tilde{H}$};
      \node (D/i) [below = of H'] {$\Lift{D_i}$};
      \draw[>->] (D/w) to node [above] {$\tilde{e}_\infty$} (H');
      \draw[->] (H') to node [right] {$\pi\Sub{i<\infty}\circ p_\infty$} (D/i);
      \draw (D/w) to node [sloped,below] {$\pi\Sub{i<\infty}$} (D/i);
    \end{tikzpicture}
  \]

  This can be seen most easily by chasing an element $\sigma\in D_\infty$:
  \begin{align*}
    \pi\Sub{i<\infty}\tilde{p}_\infty\tilde{e}_\infty\,\sigma &=
    \DLub{}\Compr{
      \pi\Sub{i<\infty}
      \tilde{p}_\infty
      \tilde{e}_j
      \pi\Sub{j<\infty}
      \sigma
    }{
      j \in \ICat \text{ s.t. } \IsDefd{\sigma_j}
    }
    \\
    &=
    \DLub{}\Compr{
      \pi\Sub{i\leq k}
      \pi\Sub{k<\infty}
      \tilde{p}_\infty
      \tilde{e}_k
      \epsilon\Sub{j\leq k}
      \pi\Sub{j<\infty}
      \sigma
    }{
      j \in \ICat \text{ s.t. } \IsDefd{\sigma_j}
    }
    && \prn{k\geq i,j}
    \\
    &=
    \DLub{}\Compr{
      \pi\Sub{i\leq k}
      \tilde{p}_k
      \tilde{e}_k
      \epsilon\Sub{j\leq k}
      \pi\Sub{j<\infty}
      \sigma
    }{
      j \in \ICat \text{ s.t. } \IsDefd{\sigma_j}
    }
    \\
    &=
    \DLub{}\Compr{
      \pi\Sub{i\leq k}
      \epsilon\Sub{j\leq k}
      \pi\Sub{j<\infty}
      \sigma
    }{
      j \in \ICat \text{ s.t. } \IsDefd{\sigma_j}
    }
    \\
    &=
    \DLub{}\Compr{
      \pi\Sub{i\leq \infty}
      \epsilon\Sub{j\leq \infty}
      \pi\Sub{j<\infty}
      \sigma
    }{
      j \in \ICat \text{ s.t. } \IsDefd{\sigma_j}
    }
    \\
    &=
    \pi\Sub{i<\infty}
    \DLub{}\Compr{
      \epsilon\Sub{j\leq \infty}
      \pi\Sub{j<\infty}
      \sigma
    }{
      j \in \ICat \text{ s.t. } \IsDefd{\sigma_j}
    }
    \\
    &=
    \pi\Sub{<\infty}\sigma
  \end{align*}
\end{node}

\begin{node}
  We must check that $e_\infty$ is a section of $p_\infty$, \ie the
  following triangle commutes:
  \[
    \begin{tikzpicture}[diagram]
      \node (L/D/w) {$\Lift{D_\infty}$};
      \node (L/H) [right = of L/D/w] {$\Lift{H}$};
      \node (L/D/w') [below = of L/H] {$\Lift{D_\infty}$};
      \draw[semb] (L/D/w) to node [above] {$e_\infty$} (L/H);
      \draw[sproj] (L/H) to node [right] {$p_\infty$} (L/D/w');
      \draw[double] (L/D/w) to (L/D/w');
    \end{tikzpicture}
  \]

  Considering the universal property of the partial map classifier, it suffices
  to check that both sides of the desired equation can be the bottom map in a
  cartesian square like the following:
  \[
    \DiagramSquare{
      west/style = >->,
      east/style = >->,
      south/style = {exists,->},
      north/style = double,
      nw/style = pullback,
      ne = D_\infty,
      se = \Lift{D_\infty},
      sw = \Lift{D_\infty},
      nw = D_\infty,
      east = \eta\Sub{D_\infty},
      west = \eta\Sub{D_\infty},
    }
  \]

  This obviously holds for the identity map, so it remains to check it for the
  upper-right composite. We employ
  \cref{node:partial:support-of-emb-infty,node:partial:tilde-emb-section}:
  \[
    \begin{tikzpicture}[diagram]
      \SpliceDiagramSquare<0/>{
        width = 2.5cm,
        nw/style = pullback,
        ne/style = pullback,
        east/style = >->,
        west/style = >->,
        south/style = semb,
        south/node/style = upright desc,
        north/node/style = upright desc,
        ne = \tilde{H},
        se = \Lift{H},
        sw = \Lift{D_\infty},
        nw = D_\infty,
        west = \eta\Sub{D_\infty},
        south = e_\infty,
        north = \tilde{e}_\infty,
      }
      \SpliceDiagramSquare<1/>{
        width = 2.5cm,
        glue = west,
        glue target = 0/,
        east/style = >->,
        south/style = sproj,
        south/node/style = upright desc,
        north/node/style = upright desc,
        ne = D_\infty,
        se = \Lift{D_\infty},
        east = \eta\Sub{D_\infty},
        north = \tilde{p}_\infty,
        south = p_\infty,
      }
      \draw[double,bend left=30] (0/nw) to (1/ne);
    \end{tikzpicture}
  \]
\end{node}

\begin{node}
  It remains to check that $\epsilon_\infty\circ p_\infty\leq
  \ArrId{\Lift{H}}$. By transitivity and the definition of $e_\infty$ as a
  least upper bound, it suffices to observe that each of the following maps is
  smaller than $\ArrId{\Lift{H}}$:
  \[
    \begin{tikzpicture}[diagram]
      \node (L/H) {$\Lift{H}$};
      \node (L/D/w) [right=of L/H] {$\Lift{D_\infty}$};
      \node (L/D/i) [right=2.5cm of L/D/w] {$\Lift{D_i}$};
      \node (L/H') [right=of L/D/i] {$\Lift{H}$};
      \draw[sproj] (L/H) to node [above] {$p_\infty$} (L/D/w);
      \draw[sproj] (L/D/w) to node [above] {$\pi\Sub{i<\infty}$} (L/D/i);
      \draw[semb] (L/D/i) to node [above] {$e_i$} (L/H');
      \draw[sproj,bend right=30] (L/H) to node [below] {$p_i$} (L/D/i);
    \end{tikzpicture}
  \]

  But this follows from the fact that each $e_i\dashv p_i$ is an ep-pair.
\end{node}

\paragraph{Uniqueness of the mediating map}

\begin{node}
  We must argue that $\SProjMor[p_\infty]{\Lift{H}}{\Lift{D_\infty}}$ is the \emph{only} projection map making
  the following diagram commute:
  \[
    \begin{tikzpicture}[diagram]
      \node (D/w) {$\brc{\Lift{D_\infty}}$};
      \node (D) [right = 2.5cm of D/w] {$\Lift{D_\bullet}$};
      \node (H) [above = of D/w] {$\brc{\Lift{H}}$};
      \draw[sproj] (D/w) to node [below] {$\pi\Sub{\bullet<\infty}$} (D);
      \draw[sproj] (H) to node [left] {$\brc{p_\infty}$} (D/w);
      \draw[sproj] (H) to node [sloped,above] {$p_\bullet$} (D);
    \end{tikzpicture}
  \]

  We therefore fix another strict projection
  $\SProjMor[q]{\Lift{H}}{\Lift{D_\infty}}$ with this property.
\end{node}

\begin{node}
  Considering the universal property of the partial map classifier, it suffices
  to check that the maps $\tilde{p}_\infty$ and $\tilde{q}$ indicated
  below are equal:
  \[
    \DiagramSquare{
      nw/style = pullback,
      west/style = >->,
      east/style = >->,
      south/style = sproj,
      nw = \tilde{H},
      sw = \Lift{H},
      ne = D_\infty,
      se = \Lift{D_\infty},
      east = \eta\Sub{D_\infty},
      south = p_\infty,
      north = \tilde{p}_\infty,
    }
    \qquad
    \DiagramSquare{
      nw/style = pullback,
      west/style = >->,
      east/style = >->,
      south/style = sproj,
      nw = \tilde{H},
      sw = \Lift{H},
      ne = D_\infty,
      se = \Lift{D_\infty},
      east = \eta\Sub{D_\infty},
      south = q,
      north = \tilde{q},
    }
  \]

  But this follows immediately from our assumptions, using the fact that maps
  into $D_\infty$ are completely determined by their behavior on projections
  $\Mor{D_\infty}{\Lift{D_i}}$.
\end{node}

\section{Other kinds of predomains}

It is worth noting that all the arguments above adapt \emph{mutatis mutandis}
to other notions of predomains characterized by closure under suprema of more
restricted kinds of directed subset; for instance, our arguments establish that
both $\omega\Con{cpo}$ and $\Con{p}\omega\Con{cpo}$ are closed under bilimits
of $\omega$-chains of embeddings.

\section{Axiomatic and synthetic domain theory}

\subsection{Kleisli models of axiomatic domain theory}

\begin{node}
  \citet{fiore-plotkin:1996} describe a simple recipe to produce models of
  \emph{axiomatic domain theory}~\citep{fiore:1994} that extend cleanly to
  sheaf models of \emph{synthetic domain
  theory}~\citep{hyland:1991,fiore-rosolini:1997} supporting recursive types.
  We recapitulate some definitions from \citet{fiore-plotkin:1996} below.
\end{node}

\begin{node}
  Let $\CCat$ be a category with an initial object and a dominance $\Sigma$, along with a
  $\Sigma$-partial map classifier monad $\Lift$.  An \emph{inductive
  fixed-point object} in $\CCat$ is defined by \opcit to be an $\Lift$-invariant object
  $\sigma:\Lift{\bar\omega} \cong \bar\omega$ together with a global element
  $\Mor[\infty]{\ObjTerm{\CCat}}{\bar{\omega}}$ such that the following conditions are satisfied
  \begin{enumerate}
    \item The object $\bar\omega$ is the colimit of the following $\omega$-chain:
      \[
        \begin{tikzpicture}[diagram]
          \node (0) {$\ObjInit{\CCat}$};
          \node (1) [right = of 0] {$\Lift{\ObjInit{\CCat}}$};
          \node (2) [right = of 1] {$\Lift^2{\ObjInit{\CCat}}$};
          \node (3) [right = of 2] {$\ldots$};
          \draw[->] (0) to node [above] {$!$} (1);
          \draw[->] (1) to node [above] {$\Lift{!}$} (2);
          \draw[->] (2) to node [above] {$\Lift^2{!}$} (3);
        \end{tikzpicture}
      \]
      (We elide the details of the cocone that we are claiming to be universal.)

    \item The following diagram commutes:
      \[
        \begin{tikzpicture}[diagram]
          \node (1) {$\ObjTerm{\CCat}$};
          \node (w) [right = 2.5cm of 1] {$\bar\omega$};
          \node (L/w) [below = 1.5cm of w] {$\Lift{\bar\omega}$};
          \node (w') [below = 1cm of L/w] {$\bar\omega$};
          \draw[->] (1) to node [above] {$\infty$} (w);
          \draw[>->] (w) to node [right] {$\eta\Sub{\bar\omega}$} (L/w);
          \draw[->] (L/w) to node [right] {$\cong$} (w');
          \draw[->] (1) to node [sloped,below] {$\infty$} (w');
        \end{tikzpicture}
      \]
  \end{enumerate}

\end{node}

\begin{node}
  A \emph{monadic base} is defined by \opcit to be a cartesian closed category $\CCat$
  with an initial object, a dominance $\Sigma$ whose lift monad is written
  $\Lift$, and an inductive fixed point object $\bar\omega$ such that the
  Eilenberg--Moore category $\CCat\Sup{\Lift}$ is closed under tensor products
  and linear homs.
\end{node}

\begin{node}
  Let $\CCat$ be a monadic base; then the Kleisli category $\CCat\Sub{\Lift}$
  is said to be a \emph{Kleisli model of axiomatic domain theory} by \opcit if
  it is $\CCat$-algebraically compact, \ie has \emph{free} algebras for
  $\CCat$-enriched endofunctors.
\end{node}

\begin{node}
  Most general results that derive algebraic compactness for $\CCat\Sub{\Lift}$
  require various assumptions that do not hold in cases of interest; for
  instance, they may use the fact that $\CCat\Sub{\Lift} = \CCat\Sup{\Lift}$,
  and they may require $\CCat$ to be locally presentable. Neither of these is
  the case for the category of dcpos in a constructive setting --- indeed, the
  $\dcpo$ will never be locally presentable, and the identification of the
  Kleisli and Eilenberg--Moore categories for the lifting monad is not
  topos-valid.
\end{node}

\begin{node}
  \citet{fiore-plotkin:1996} describe a simple recipe to verify the assumptions
  required by Kleisli models of axiomatic domain theory, quoting a lecture of
  Plotkin from 1995:
  \begin{quote}
    If the Kleisli category $\CCat\Sub{\Lift}$ has an enriched zero object and
    bilimits of $\omega$-chains of embedding-projection pairs, then it is
    $\CCat$-algebraically compact.
  \end{quote}
\end{node}

\begin{node}
  Therefore the results contained in this note establish that a Kleisli model
  of axiomatic domain theory can be fashioned from the dcpos of an arbitrary
  Grothendieck topos. In particular, for a topos $\ECat$ we may consider the
  (external) category $\dcpo_\ECat$ of global dcpo-objects in $\ECat$ and
  continuous maps between them. We have a dominance $\Sigma$ in $\dcpo_\ECat$
  induced by the universal open inclusion $\Mor|>->|{\ObjTerm}{\Omega}$, where
  $\Omega$ is the Sierpi\'nski domain corresponding to the subobject classifier
  of $\ECat$. The corresponding lifting monad for $\Sigma$ has as a
  carrier-object the partial map classifier monad of $\ECat$. It is easy to see
  that the Kleisli category $\pdcpo_\ECat = \prn{\dcpo_\ECat}\Sub{\Lift}$ has
  an enriched zero object, and in this note we have proved that it is closed
  under bilimits $\omega$-chains of embedding-projection pairs --- indeed, we
  have proved the stronger result that it is closed under internal bilimits of
  arbitrary directed posetal diagrams.
\end{node}

\subsection{The Fiore--Plotkin conservative extension result}

\begin{node}
  One of the main results of \citet{fiore-plotkin:1996} is to establish that
  every \emph{small} Kleisli model of axiomatic domain theory extends to a
  sheaf model of an infinitary version of synthetic domain theory (SDT) in which the
  \emph{well-complete} objects serve as a category of predomains that contains
  $\CCat$ in a fully faithful way.
\end{node}

\begin{node}
  This sheaf model of synthetic domain theory is obtained in a very simple way;
  one considers the closed subtopos of $\tilde{\CCat}\subseteq \PrTop{\CCat}$
  determined by the subterminal object $\Yo{\ObjInit{\CCat}}$; concretely, when
  $\ObjInit{\CCat}$ is strict, this amounts to presheaves taking
  $\ObjInit{\CCat}$ to the terminal object. The lifting monad in
  $\tilde{\CCat}$ is obtained by Yoneda extension, and therefore restricts to
  the original lifting monad.
\end{node}

\begin{node}
  The inductive fixed point object $\bar\omega\in\CCat$ becomes the final
  coalgebra for the lift monad in $\tilde{\CCat}$; the \emph{initial} algebra
  for the lift monad is denoted $\omega$, and comes equipped with a canonical
  monomorphism $\Mor|>->|{\omega}{\bar\omega}$. Then a \emph{complete} object
  is one from whose perspective the lift algebra and the final lift coalgebra
  appear to coincide, \ie it is internally orthogonal to
  $\Mor|>->|{\omega}{\bar\omega}$. A \emph{well-complete} object is an object
  whose lift is complete. The well-complete objects form a category of
  predomains, and the Yoneda embedding $\EmbMor{\CCat}{\tilde{\CCat}}$ factors
  through the inclusion of well-complete objects into $\tilde{\CCat}$.  Among
  the well-complete objects, the $\Lift$-algebras support recursion by means of
  their orthogonality principle; moreover, domain equations can be solved.
\end{node}

\begin{node}
  By restricting $\dcpo_\ECat$ to small subcategory (\eg by replacing $\ECat$
  with the full internal subcategory determined by a topos universe in the
  sense of \citet{streicher:2005}), we may therefore compose the results of
  \citet{fiore-plotkin:1996} to obtain models of synthetic domain theory based
  on ``exotic'' concrete domain theories; for instance, if $\ECat$ is the topos
  of sheaves on a topological space $X$, the domain theory over $\ECat$
  includes partial maps that terminate only over certain regions of $X$.
\end{node}

\printbibliography

\end{document}